\documentstyle[PASJadd,epsf]{PASJ95}

\newcommand{\vol}{53}
\newcommand{\no}{3}
\newcommand{\lett}{}
\newcommand{\spage}{0}
\newcommand{\rdate}{2000 December 19}
\newcommand{\adate}{2001 March 18}

\begin{document}
\title{{\it JHK} Spectra of the $z=2.39$ Radio Galaxy 53W002 $^{1}$}

\author{
Kentaro {\sc Motohara}$^{2,3}$\\
Toru {\sc Yamada}$^{4,6}$, Fumihide {\sc Iwamuro}$^{3}$, Ryuji {\sc
Hata}$^{3}$, Tomoyuki {\sc Taguchi}$^{3}$, Takashi {\sc
Harashima}$^{3}$,\\
Toshinori {\sc Maihara}$^{3,5}$, Masanori {\sc Iye}$^{4}$, Chris {\sc
Simpson}$^{2}$, and Michitoshi {\sc Yoshida}$^{4,7}$ 
 \\[12pt]
 $^{2}$ {\it Subaru Telescope, National Astronomical Observatory of
 Japan,}\\
 {\it 650 North A'ohoku Place, Hilo, HI 96720, USA} \\
 {\it motohara@naoj.org}  \\
 $^{3}$ {\it Department of Physics, Kyoto University, Kitashirakawa, Kyoto
606-8502} \\
 $^{4}$ {\it Optical and Infrared Astronomy Division, National Astronomical
Observatory of Japan, Mitaka, Tokyo 181-8588}\\
 $^{5}$ {\it Department of Astronomy, Kyoto University, Kitashirakawa, Kyoto
606-8502} \\
 $^{6}$ {\it Astronomical Institute, Tohoku University, Aoba-ku, Sendai, Miyagi 980-8578} \\
 $^{7}$ {\it Okayama Astrophysical Observatory, National Astronomical
Observatory of Japan,}\\
{\it Kamogata-cho, Asakuchi-gun, Okayama 719-0232}\\
}

\abst{We present low-resolution, near-IR {\it JHK} spectra of
the weak $z=2.39$ radio galaxy 53W002, obtained with the OH-airglow
Suppressor spectrograph (OHS) and Cooled Infrared Spectrograph and
Camera for OHS (CISCO) on the Subaru Telescope.  They cover rest-frame
wavelengths of 3400--7200 \AA, and the emission lines of [O\,{\sc
ii}]\,$\lambda3727$, H$\beta$, [O\,{\sc iii}]\,$\lambda\lambda4959, 5007$,
H$\alpha$, [N\,{\sc ii}]\,$\lambda\lambda6548, 6583$ and [S\,{\sc
ii}]\,$\lambda\lambda6716, 6731$ were detected. Using the $\rm
H\alpha/H\beta$ line ratio, we find an extinction of $E(B-V)=0.14$.
The emission-line ratios are reproduced by a cloud of electron density
$n_{\rm e}=1\times10^{3-4}$ $\rm (cm^{-3})$ with solar metallicity,
ionized by an $\alpha=-0.7$ power-law continuum with ionizing
parameter $U=1\times10^{-3}$.  In addition to these emission lines, we
make the first spectroscopic confirmation of the Balmer discontinuity
in a high-$z$ radio galaxy. Together with rest-frame UV photometry from
the literature, we show that at least 1/3 of the present stellar mass was
formed in the current starburst. 
The stellar mass was estimated to be $(1-1.4)\times10^{11}\,M_{\odot}$ by
one-component model fitting, 
which is smaller than that of typical $z\sim1$ B2/6C radio galaxies. 
We suggest that 53W002 is
currently assembling a large part of its stellar mass through merger
events with the surrounding sub-galactic clumps, some of which can be
identified with the L$\alpha$ emitters detected in narrow-band
imaging. After a few such events over the next few Gyr, 53W002 will
evolve into a massive elliptical galaxy.}

\kword{galaxies: active --- galaxies: individual (53W002) --- galaxies: spectroscopy}

\maketitle

\footnotetext[1]{
Based on data collected at Subaru Telescope, which is
operated by the National Astronomical Observatory of Japan.
}

\thispagestyle{headings}

\section{Introduction}
Many high-$z$ field galaxies were discovered in the 1990s using
the UV drop-out technique.  However, these objects are rather faint,
and infrared spectroscopy is difficult (Pettini et al. 1998).  On the
other hand, high-$z$ radio galaxies are relatively bright, and their
properties have been fairly well-studied. In addition, as the progenitors of
present-day massive galaxies, they are important for studying galaxy
formation. 

53W002 is a $z=2.39$ galaxy with relatively weak radio power ($S_{1.4
\rm GHz}\sim50 \rm \,mJy$; Windhorst et al.\ 1991), indicating that its AGN activity is not
extreme and it could therefore be suitable for a study of the
underlying stellar population. Broad-band photometry (Windhorst et al.\ 1991) reveals a
possible break at $\lambda_{\rm rest} \sim 4000$\,\AA, and suggests
recent star-formation activity from spectral model fitting.  
Its rest-frame UV spectrum suggests
that the main-sequence turnoff age of the current starburst is
0.25--0.3 Myr (Windhorst et al.\ 1991).  Radio observations have
detected lines from CO molecules concentrated at the center of the
galaxy (Yamada et al.\ 1995, Scoville et al.\ 1997, 
Alloin et al.\ 2000), which may be other indications of star-formation
activity.

The radial surface brightness profile of 53W002, as revealed by HST, is
well-modeled by a de Vaucouleurs profile, suggesting that the galaxy
is already dynamically relaxed. A $V-I$ radial color gradient suggests
that the center of the galaxy consists of a young population (0.3--0.5
Gyr), while older stars (0.5--1.0 Gyr) exist in the outer ($r>10$ kpc)
region (Windhorst et al.\ 1994).

Previous infrared spectroscopy has revealed the presence of the
H$\alpha$+[N\,{\sc ii}] lines in the $K$-band (Eales, Rawlings
1993), but a detailed discussion was precluded by their limited
sensitivity. We have therefore carried out low-resolution infrared
spectroscopy of 53W002 to investigate the properties of its
emission-line clouds and underlying stellar population.  For {\it
JH\/}-band spectroscopy, we used a newly-commissioned instrument of
the Subaru telescope, an OH-airglow suppression Spectrograph (OHS;
Maihara et al.\ 1994, Iwamuro et al.\ 2001).  OHS is a spectroscopic
filter which eliminates almost all of the OH airglow lines in the {\it
JH\/}-band, which are the major source of atmospheric background
radiation at near-infrared wavelengths below $2\, \mu \rm m$. Using OHS, the
background level is suppressed by 96\% and the sensitivity becomes 1
magnitude fainter. The output from OHS is an undispersed beam from
which OH airglow is suppressed; this is then sent to the Cooled
Infrared Camera and Spectrograph for OHS (CISCO; Motohara et al.\
1998) to acquire either images or spectra. CISCO can also be used as
an ordinary infrared camera and spectrograph attached on either the
Cassegrain or the Nasmyth focus of the telescope. For $K$-band
spectroscopy, we used this mode.

We describe the observations and data reduction in section 2 and
present the results in section 3. Discussions are given out in
section 4 and we summarize our results in section 5. We adopt $H_0=65
\rm\, km\, s^{-1}$ and $q_0=0.1 $ throughout this paper. The scale at
$z=2.39$ is thus 8.5 kpc arcsec$^{-1}$.

\section{Observations and Data Reductions}

\subsection{$K$-Band Spectroscopy}

$K$-band spectroscopy of 53W002 was carried out on 1999 May 10, with
CISCO mounted at the Cassegrain focus of the Subaru telescope.  The
pixel scale was $0.\!\!^{\prime\prime}116$/pixel. A slit width of
$0.\!\!^{\prime\prime}7$ provided a wavelength resolution of $\sim430$
at $2.2\, \mu \rm m$.  The position angle of the slit was set to $145^{\circ}$
to include object 13 of Pascarelle et al.\ (1998) in the slit.
Individual exposures were 200 s, and the telescope was nodded slightly
($\sim 5^{\prime\prime}$) after every six frames to provide sky
subtraction. A total of 18 frames were taken for a total exposure time
of 3600 s. The seeing size was $0.\!\!^{\prime\prime}4$.  After the
observation, the F0 star SAO 30082 was observed for correction of the
atmospheric absorption.

\subsection{$H$-Band Imaging and the JH-Band Spectroscopy}

$H$-band imaging was carried out on 2000 May 20 and {\it JH\/}-band
spectroscopy on 2000 May 20 and 23, with OHS and CISCO mounted at the
Nasmyth focus of the Subaru telescope.  The pixel scale was
$0.\!\!^{\prime\prime}105$/pixel.

The $H$-band images were taken with OHS in the imaging mode (see Iwamuro
et al.\ 2001). The single exposure time was 50 s and the telescope was
nodded every 3 frames by $10^{\prime\prime}$ in the east--west
direction.  A total of 12 frames were acquired, for an exposure time
of 600 s.

The {\it JH\/}-band spectra were acquired using a
$0.\!\!^{\prime\prime}95$ slit, providing a resolution of $\sim210$ at
1.65$\mu$m. A position angle of $90^\circ$ was used, which was aligned
with the major axes of the radio and L$\alpha$ emission. After each
1000 s exposure, the telescope was nodded by $10^{\prime\prime}$ along
the slit. We took 4 frames each day, and the total exposure time
was 8000 s. The seeing size was $0.\!\!^{\prime\prime}6$.  The F8 star
SAO 30361 and F5 star SAO 30245 were observed after the target on May
20 and 23, respectively, for correcting the atmospheric absorption.

\subsection{Data Reduction}

The $H$-band frames were reduced though standard procedures of
flat-fielding, sky subtraction, correction of bad pixels, residual sky
subtraction, before registration and coaddition. Photometry was
performed in a $1.\!\!^{\prime\prime}0\times2.\!\!^{\prime\prime}1$
aperture at a position angle of $90^\circ$, which is equal to the
spectroscopic extraction aperture position. The flux in the $H$-band
was $11.9\pm0.7\,\mu$Jy, calibrated from an observation of FS 23.

The spectroscopic data were reduced in the standard manner of
flat-fielding, sky subtraction, correction of bad pixels and residual
sky subtraction. The spectra of the SAO F stars were used to correct
for atmospheric extinction.

The wavelength scale of the $K$-band spectrum was calibrated from
airglow lines in the raw frames. Such a procedure was impossible for
the {\it JH\/} spectrum, since all the strong OH-lines were suppressed
by OHS. We therefore calibrated our spectrum from another observation
of CISCO without OHS, where OH lines were available. We checked the
stability of the wavelength scale from all of the previous CISCO
observations, and found the systematic error to be less than 0.5
pixels ($<5 \, \rm \AA$).

Because the seeing size was less than the slit width for both
spectroscopic observations, we assumed the error in the wavelength
calibration to be 1/4 of the slit width, i.e., $\sim\pm13\, \rm \AA$ (1.5
pixels) for the $K$-band and $\sim\pm22 \, \rm \AA$ (2.4 pixels) for the {\it
JH\/}-band.  Thus, the systematic error in redshifts are 0.0006 in the
$K$-band and 0.0015 in the {\it JH\/}-band.

\section{Results}

The final {\it JHK\/} spectra are shown in figure 1. The {\it JH\/}-band
spectrum was calibrated from the $H$-band photometry and the
$K$-band spectrum from the $K'$-band photometry of Yamada et al.\ (in
preparation). A $1.\!\!^{\prime\prime}0\times2.\!\!^{\prime\prime}1$
aperture, aligned with the position angle used for the spectroscopy,
was used in both cases.

\subsection{Emission Lines}

We detected emission lines of [O\,{\sc ii}]\,$\lambda3727$, H$\beta$,
[O\,{\sc iii}]\,$\lambda\lambda4959, 5007$, H$\alpha$+[N\,{\sc
ii}]\,$\lambda\lambda6548, 6583$ and [S\,{\sc
ii}]\,$\lambda\lambda6716,6731$. [Ne\,{\sc iii}]\,$\lambda$3869 may be
present, but this region is contaminated with the edge of the Balmer
discontinuity. None of the emission lines was resolved at the
wavelength resolutions of OHS/CISCO, which were 700 km s$^{-1}$ in the
$K$-band, and 1500 km s$^{-1}$ in the {\it JH\/}-band.

Due to the low wavelength resolution, the H$\alpha$ and [N\,{\sc ii}]
lines were blended, as was the [S\,{\sc ii}] doublet. We
deconvolved these lines by fitting multiple Gaussian profiles. For
H$\alpha$+[N\,{\sc ii}], we assumed [N\,{\sc
ii}]\,$\lambda$6583/[N\,{\sc ii}]\,$\lambda$6548 = 3.0 and demanded that all
three lines have the same width. For the [S\,{\sc ii}] doublet, we
fixed the line widths to be equal to results from the
H$\alpha$+[N\,{\sc ii}] fitting. The results are given in figure 2.
The derived line ratios were [N\,{\sc
ii}]\,$\lambda$6583/H$\alpha = 2.3\pm0.6$ and [S\,{\sc
ii}]\,$\lambda6716$/[S\,{\sc ii}]\,$\lambda6731 = 0.61\pm0.44$
($2\sigma$). Other lines were fitted by single Gaussian profiles to
obtain their redshifts and to determine the continuum level; we assumed a
common redshift and line widths for [O\,{\sc ii}], H$\beta$, and
[O\,{\sc iii}]. The emission-line properties are listed in table
1. The fluxes of [O\,{\sc ii}], H$\beta$, and [O\,{\sc i}] are the value
above the continuum level.
The fluxes of the other lines are determined from
the results of Gaussian fitting, because they are blended.
The redshift difference between lines in the {\it JH\/}- and
$K$-bands might arise from a systematic error in the wavelength
calibration.

The contribution to the broad-band fluxes from the emission lines is
very large, amounting to 22\% in the $J$-band, 30\% in the $H$ and
42\% in the $K^{\prime}$-band.

We show the spatial intensity profiles of the continuum and emission
lines along the slit in figure 3.  The profiles of both [O\,{\sc iii}]
and H$\alpha$+[N\,{\sc ii}] are clearly more extended than the
adjacent continua. We estimate of the spatial extent of the line
emission to be roughly
$0.\!\!^{\prime\prime}7$--$0.\!\!^{\prime\prime}8$, or 6--7 kpc.

\subsection{Detection of Balmer Discontinuity}

The high sensitivity of OHS allows the continuum of 53W002 to be
clearly detected. We show the SED of the continuum in figure 4 with a
solid line. We removed all of the emission-lines, except for [Ne\,{\sc iii}],
using the results of the Gaussian profile fitting, and then smoothed
the resulting spectrum with a 0.05 $\mu$m boxcar filter.

A clear break in the continuum is seen at 1.3\,$\mu$m, which is the
first spectroscopic detection of the Balmer discontinuity in a
high-redshift radio galaxy. The [Ne\,{\sc iii}] emission line, which
we disregarded above, is located here, and may have affected the
position of this break. To check the effect of this emission line, we
subtract a Gaussian at the position of the line, assuming [Ne\,{\sc
iii}]\,$\lambda$3869/[O\,{\sc ii}]\,$\lambda3727=0.2$ which was the result
of simulations from CLOUDY94 carried out in the following
analysis. The continuum thus derived is displayed with a dotted line
in figure 4, and shows a slight shift in the position of the break.

\section{Discussion}

\subsection{Nature of Emission-Line Clouds}

We measure H$\alpha$/H$\beta = 3.4$. If we adopt an intrinsic value of
3.1 (Osterbrock 1989) and the SMC extinction curve of Pei (1992) with
$A_V/E(B-V)=3.1$, we derive $E(B-V)=0.14$. The extinction-corrected
emission line fluxes are listed in the fifth column of table 1.

The sulfur line ratio is a good indicator of the electron density
(Osterbrock 1989), and our measured value of [S\,{\sc
ii}]\,$\lambda6716$/[S\,{\sc ii}]\,$\lambda6731=0.61\pm0.44$ implies
$n_{\rm e}\sim1\times10^{3}\,\rm cm^{-3}$. The $2\sigma$ lower limit
is $n_{\rm e}\sim5\times10^{2}\,\rm cm^{-3}$, but we cannot constrain
the upper limit.

The L$\alpha$ flux is $2.8\times10^{-19}$ W m$^{-2}$ (Keel et
al.\ 1999). Because the L$\alpha$ morphology is smaller than our slit
size of $1.\!\!^{\prime\prime}0$ (Windhorst et al.\ 1998), we estimate
the $L({\rm L}\alpha)/L({\rm H}\beta)$ ratio to be 3.2. The
correction for the Galactic extinction of $E(B-V)=0.03$ (Burstein,
 Heiles 1982) produces $L({\rm L}\alpha)/L({\rm H}\beta)=3.5$. A
further correction for the {\em in situ\/} extinction of $E(B-V)=0.14$
results in L$\alpha$/H$\beta=17$, which is consistent, given the
uncertainty in the extinction, with a low-density value of 23
(Ferland, Osterbrock 1985).

In figure 5 we present two emission-line ratio diagrams. One shows
[N\,{\sc ii}]\,$\lambda$6583/H$\alpha$ versus [O\,{\sc
iii}]\,$\lambda$5007/H$\beta$ and the other [O\,{\sc
ii}]\,$\lambda$3727/[O\,{\sc iii}]\,$\lambda$5007 versus [O\,{\sc
iii}]\,$\lambda$5007/H$\beta$. We plot the location of 53W002 together
with the ratios of narrow-line AGNs; H\,{\sc ii} regions taken from
the literature. The radio galaxy lies in the region occupied by narrow
line AGNs and the emission-line cloud is therefore ionized by a hidden
nucleus.

To investigate the physical state of the emission-line cloud, we
carried out photoionization calculations using the code CLOUDY94
(Ferland 1999, 2000). We assume a spectral index of $\alpha=-0.7$ for
the ionizing continuum, taken from the composite spectra of radio-loud
QSOs (Cristiani, Vio 1990; Zheng et al.\ 1997). The observed line
ratios are reproduced with $Z\sim Z_{\odot}$, $\log U\sim-3$, and
$n_{\rm e}=1\times10^{3-4}\,\rm cm^{-3}$ (see solid and dashed lines
in figure 5). We could not reproduce the observed ratios with an
$\alpha=-1.5$ power law continuum (dot-dashed lines in figure 5), nor
by models with $0.1\, Z_{\odot}$ metallicity (dashed-long-dashed lines
in figure 5).

We then estimated the mass of ionized gas from the physical values
derived above and the H$\alpha$ line luminosity, using the following
equations:
\begin{eqnarray}
 L_{\rm H\alpha}&=&n_{\rm e}^2 \alpha_{\rm H\alpha}^{\rm eff}
h \nu_{\rm H\alpha} V f, \\
 M_{\rm gas} & = & V f n_{\rm e} m_{\rm H},
\end{eqnarray}
where $h$ is the Planck constant, $\nu_{\rm H\alpha}$ the frequency of
the H$\alpha$ line, $V$ the volume of the line-emitting cloud, $f$ the
filling factor, $m_{\rm H}$ the hydrogen mass, and $\alpha_{\rm
H\alpha}^{\rm eff}=6.04\times10^{-14} \rm(cm^3\,s^{-1})$ (Osterbrock
1989) the H$\alpha$ recombination coefficient under Case B. We derive
a gas mass of $10^{7-8}\,M_\odot$.

A rough estimate of the filling factor, $f$, of the cloud can be made by
assuming the cloud to be a sphere of 3 kpc radius (section 3.1). We
calculate $f \sim 10^{-5}$--$10^{-7}$, consistent with the typical
values of low-$z$ radio galaxies of $10^{-4}$--$10^{-6}$ (Heckman et al.\
1982; van Breugel et al.\ 1985).

High-resolution HST imaging shows the presence of an extended blue
cloud which peaks $0.\!\!^{\prime\prime}45$ west of the nucleus
(Windhorst et al.\ 1998). Our emission-line region, however, almost
coincides with the continuum core. Therefore, what we are seeing is
not those blue clouds, but a cloud which envelops the whole
galaxy.  This gas could be a remnant of a previous merger event,
resulting in the starburst activity observed in our spectra as the Balmer
discontinuity (see section 4.2.2).  Metallicity as high as
$Z_\odot$ might be produced during this period.

\subsection{SED and Age of the Host Galaxy}

The spectral energy distribution of 53W002 is shown by the open circles
and squares in figure 6. The optical data were taken from Windhorst et
al.\ (1991) and the infrared data from line-free regions of our
spectra. An additional data point was derived from the continuum after
subtracting the [Ne {\sc iii}] line assuming [Ne {\sc iii}]/[O {\sc
ii}]\ =\ 0.2$\pm$0.1.  Because Windhorst et al.\ (1991) measured the total
magnitudes, our infrared data were multiplied by a factor of 1.2 to
match their $H$- and $K$-band photometry. We disregard the $U$-band
data of Windhorst et al.\ (1991), which is shortward of L$\alpha$ in
the rest-frame, and is likely to be strongly attenuated by the
L$\alpha$ forest from the surrounding objects.

Next, we removed two non-stellar components from the SED. One is the
nebular continuum emission and the other is the scattered light from
the hidden nucleus.

The strength of the nebular continuum was calculated from the H$\beta$
emission line flux using the tabulated coefficients of Aller (1984)
for $T_{\rm e}=1\times10^{4}$\,K. We ignore the contribution from He
recombination, which our CLOUDY94 simulations indicate will be weak.

We assume the intrinsic shape of the continuum from the hidden nucleus
to be an $\alpha=-0.7$ power law. Since we do not know whether the
scatterers are dust grains or electrons, we consider both cases.
Because electron-scattering is achromatic, the scattered spectrum is the
same as the incident spectrum, while the dust-scattered spectrum was
from Cimatti et al.\ (1994). After applying the extinction of
$E(B-V)=0.14$, the flux was scaled to match the upper limits of the
scattered AGN contribution estimated by Windhorst et al.\ (1998) as
shown in figure 6.

The resulting SEDs of the underlying stellar population, after
subtracting these two components, is shown by the filled circles and
triangles in figure 6. We tried to fit these SEDs with a few models of
galactic spectra.

\subsubsection{One-component model}

First, we fitted these SEDs with single-component synthetic spectra
produced by the spectrophotometric galaxy evolution code PEGASE (Fioc,
 Rocca-Volmerange 1997), to which an extinction of $E(B-V)=0.14$ was
applied. Instead of the SMC extinction curve, we applied the
extinction curve of Calzetti et al.\ (1994), which is an empirical
relation for the stellar UV continuum derived from the spectra of
nearby starburst regions. We considered three star-formation histories
--- an instantaneous burst, and two exponential bursts with
$e$-folding times of 200 Myr and 1 Gyr --- and the age was left as a
free parameter. We adopted the Scalo IMF. We also fit our models
to the observed SED, to investigate the effect of subtracting the
non-stellar emission. These results are given in table 2.

The results are not very sensitive to the assumed non-stellar
contributions, since the age of the galaxy is mainly determined by the
strength of the Balmer discontinuity. Although the fitted parameters
vary according to the assumed star-formation history, the age is
always young: 70--700 Myr, corresponding to a formation redshift
$z_{\rm form}=2.5$--3.2.

Windhorst et al.\ (1991) showed, from the far-UV SED and possible
detections of UV absorption lines, that the main-sequence cutoff age
is 250--320 Myr, so we adopt the $\tau=200$ Myr model in the following
discussions. The stellar mass of the galaxy is then $\sim
10^{11}\,M_\odot$, which is of the order of that of an $L^*$ galaxy.

\subsubsection{Two-component model}

According to the high-resolution HST imaging, 53W002 has a radial
color gradient, suggesting that the outskirts of the galaxy
is comprised of older stars (0.5--1 Gyr; Windhorst et al.\ 1994). There
may therefore have been additional star-formation episodes prior to
the current one, and we attempt to fit the SED with two stellar
components to evaluate how much stellar mass was created by the
current starburst activity.

The models consisted of an old component of fixed age and a younger
component of unknown age. A $\tau=200$ Myr exponential burst with
Scalo IMF was assumed for both components. We varied the fraction of
the initial gas mass in the old component from 0.1 to 0.9 in 0.1
steps. Two ages for the old component were assumed: 
one is 1 Gyr, the age suggested by Windhorst et al. (1994); the other
is 3 Gyr, the maximum age allowed by the age of the universe.
The results of the model fitting are shown in table 3 and
figure 7.

The $\chi^2$-test was used to assess the results of the fitting.  We
reject models which are not consistent with the data at the 99\%
confidence level, which corresponds to $\chi^2>25$. We therefore
exclude models where the mass fraction of old component is 0.9 or
larger. We also reject models which produce an age of less than 200
Myr for the current burst, which is a lower limit derived from the UV
spectrum (Windhorst et al.\ 1991), and only those models whose old
mass fraction is less than 0.7 remain. We therefore conclude that
53W002 is undergoing a massive starburst, and that at least 1/3 of its
stellar mass is forming in the current burst.

\subsubsection{Mass and formation of 53W002}
We next compare the stellar mass with those of other radio galaxies in the literatures.
Because these mass estimations were carried out by single-component SED
fitting, we use the mass of 53W002 derived using the one-component model.

The stellar mass of the galaxy was estimated to be less than
$1.4\times10^{11}\,M_\odot$ by the one-component model, similar to that of $z=1.824$ radio galaxy
3C 256 (Simpson et al.\ 1999). This galaxy also underwent a recent
starburst and its stellar mass is only $1.2\times10^{11}\,M_\odot$
(converted to our adopted cosmology). Best et al.\ (1998) estimated
the masses of 26 $z\sim1$ 3C radio galaxies using the GISSEL spectral
synthesis codes of Bruzual and Charlot (1993), and derived a median value of
$4.4\pm1.7\times10^{11}\,M_\odot$ (again, converted to our adopted cosmology).
We determine that the effect of the different IMFs used is only 10\%,
and therefore conclude that the mass of 53W002 is less than 1/3 of
a typical $z\sim 1$ 3C radio galaxy. However, Eales et al.\ (1997)
reported a correlation between the radio and $K$-band luminosities of
$z\sim1$ radio galaxies, and the radio luminosity of 53W002 is similar
to those of the B2/6C radio galaxies, which are on average $\sim 0.6$
mag fainter than the 3C objects. If we assume that the stellar mass
correlates with the $K$-band luminosity, the mass of $z\sim1$ B2/6C
radio galaxies is $\sim 2.5\times10^{11}\,M_{\odot}$, which is still
twice the value of 53W002.

There are 17 candidate L$\alpha$-emitters in the field of 53W002, and
7 of them have been spectroscopically confirmed to lie at the same
redshift as the radio galaxy (Pascarelle et al.\ 1996a,b,1998; Keel et al.\ 1999). These objects are of sub-galactic
size, and were postulated to be `building blocks' which will merge
into a luminous galaxy by the present epoch. If 53W002 experiences
additional starburst activity every 500 Myr by merging with these
surrounding sub-galactic clumps, it will grow into a massive galaxy of
a few$\times10^{11}\,M_\odot$ by $z=1$. These results strongly support
the hypothesis that 53W002 is a forming massive elliptical galaxy
undergoing multiple merger events, and an ancestor of present-day
luminous galaxies.

\section{Summary}

We have performed infrared {\it JHK\/} spectroscopy of the weak radio
galaxy 53W002 at $z=2.39$, using CISCO and the newly commissioned
instrument OHS on the Subaru telescope.

We detected the rest-frame optical emission lines of [O\,{\sc
ii}]\,$\lambda3727$, H$\beta$, [O\,{\sc iii}]\,$\lambda\lambda4959, 5007$,
H$\alpha$, [N\,{\sc ii}]\,$\lambda\lambda6548, 6583$ and [S\,{\sc
ii}]\,$\lambda\lambda6716, 6731$. These emission lines were reproduced
by a model cloud with $Z\sim Z_\odot$, $\log U\sim -3$, and $n_{\rm
e}=1\times10^{3-4}\,\rm(cm^{-3})$, ionized by an $\alpha=-0.7$
power-law continuum. The mass of ionized gas in the cloud was
estimated to be $M_{\rm gas}=10^{7-8}\,M_\odot$, and the filling
factor was calculated to be as small as $10^{-5}$--$10^{-7}$,
indicating significant clumpiness. We found no offset between the
peak positions of the line and continuum emission, which suggests that
the emission line cloud is not the extended blue cloud reported in
Windhorst et al.\ (1998).

The Balmer discontinuity was detected clearly for the first time in
the continuum of a high-redshift radio galaxy. We were able to fit the
rest-frame UV--optical continuum by a 70--700 Myr-old stellar
population. A fit with two-component model indicated that at least
one-third of the total stellar mass is involved in the current starburst. We
estimated the total stellar mass to be 1--$1.4\times10^{11}\,M_\odot$ by
one-component model,
which is smaller than typical $z\sim1$ B2/6C radio galaxies. 
The mass becomes twice as much if we introduce the two-component model, 
but is still smaller than that of $z\sim1$ 3C radio galaxies.
We therefore conclude that 53W002 is now assembling a large part of its stellar
mass through mergers with the surrounding sub-galactic objects, and
will end up as a massive elliptical observed today.
\\[6pt]

We are indebted to all staff members of the Subaru telescope, NAOJ.  We
appreciate M.\ Fioc and B.\ Rocca-Volmerange for generously offering
their galaxy modeling code, PEGASE and G.\ Ferland for the spectral
synthesis code CLOUDY94.

\section*{References}
\small
\re
Aller, L. H.\ 1984, Physics of Thermal Gaseous Nebulae (Dordrecht: Reidel),
p.98
\re
Alloin, D., Barvainis, R., \& Guilloteau, S.\ 2000, ApJ, 528, L81
\re
Best, P. N., Longair, M. S., \& R\"ottgering, H. J. A.\ 1998, MNRAS, 295, 549
\re
Burstein, D., \& Heiles, C.\ 1982, AJ, 87, 1165
\re
Buzual, G., \& Charlot, S.\ 1993, ApJ, 405, 538
\re
Calzetti, D., Kinney, A. L., \& Storchi-Bergmann, T.\ 1994, ApJ, 429, 582
\re
Cimatti, A., di Serego Alighieri, S., Field, G. B., \& Fosbury, R. A. E.\ 1994, ApJ, 422, 562
\re
Costero, R., \& Osterbrock, D. E.\ 1977, ApJ, 211, 675
\re
Cristiani, S., \& Vio, R.\ 1990, A\&A, 227, 385
\re
Eales, S. A., \& Rawlings, S.\ 1993, ApJ, 411, 67
\re
Eales, S. A., Rawlings, S., Law-Green, D., Cotter, G., \& Lacy, M.\ 1997, MNRAS, 291, 593
\re
Fioc, M., \& Rocca-Volmerange, B.\ 1997, A\&A, 326, 950
\re
Ferland, G. J.\ 2000, Rev. Mex. Astron. Astrofis., 9, 153
\re
Ferland, G. J.\ 1999, in ASP Conf. Ser. 216, Astronomical Data
Analysis Software and Systems IX, ed. N. Manset, C. Veillet, \&
D. Crabtree (San Francisco: ASP), 32 
\re
Ferland, G. J., \& Osterbrock, D. E.\ 1985, ApJ, 289, 105
\re
French, H. B.\ 1980, ApJ, 240, 41
\re
Heckman, T. M., Miley, G. K., Balick, B., van Breugel, W. J. M., \&
Butcher, H. R.\ 1982, ApJ, 262, 529
\re
Iwamuro, F., Motohara, K., Maihara, T., Hata, R., \& Harashima, T.\
2001, PASJ, 53, 355
\re
Keel, W. C., Cohen, S. H., Windhorst, R. A., \& Waddington, I.\ 1999, AJ, 118, 2547
\re
Koski, A. T.\ 1978, ApJ, 223, 56
\re
Maihara, T., Iwamuro, F., Oya, S., Tsukamoto, H., Hall, D. N., Cowie,
L. L., Tokunaga, A. T., \& Pickles, A. J.\ 1994, Proc.\ SPIE, 2198, 194
\re
McCall, M. L., Rybski, P. M., \& Shields, G. A.\ 1985, ApJS, 57, 1
\re
Motohara, K., Maihara, T., Iwamuro, F., Oya, S., Imanishi, M., Terada, H.,
Goto, M., Iwai, J., et al.\ 1998, Proc.\ SPIE, 3354, 659 
\re
Osterbrock, D. E.\ 1989, Astrophysics of Gaseous Nebulae and Active
Galactic Nuclei (Mill Valley: University Science Books)
\re
Pascarelle, S. M., Windhorst, R. A., Driver, S. P., Ostrander, E. J., \&
Keel, W. C.\ 1996a, ApJ, 456, L21
\re
Pascarelle, S. M., Windhorst, R. A., Keel, W. C., \& Odewahn, S. C.\ 1996b,
Nature, 383, 45
\re
Pascarelle, S. M., Windhorst, R. A., \& Keel, W.C.\ 1998, AJ, 116, 2659
\re
Pei, Y. C.\ 1992, ApJ, 395, 130
\re
Pettini, M., Kellogg, M., Steidel, C. C., Dickinson, M., Adelberger,
K. L., \& Giavalisco, M.\ 1998, ApJ, 508, 539
\re
Simpson, C., Eisenhardt, P., Armus, L., Chokshi, A., Dickinson, M.,
Djorgovski, S. G., Elston, R., Jannuzi, B. T., et al.\ 1999, ApJ, 525, 659
\re
Scoville, N. Z., Yun, M. S., Windhorst, R. A., Keel, W. C., \& Armus, L.\
1997, ApJ, 485, L21
\re
van Breugel, W. J. M., Miley, G., Heckman, T., Butcher, H., \& Bridle,
A.\ 1985, ApJ, 290, 496
\re
Veilleux, S., \& Osterbrock, D. E.\ 1987, ApJS, 63, 295
\re
Windhorst, R. A., Burstein, D., Mathis, D. F., Neuschaefer, L. W., Bertola, F.,
Buson, L. M., Koo, D. C., Matthews, K., Barthel, P. D., \& Chambers,
K. C.\ 1991, ApJ, 380, 362
\re
Windhorst, R. A., Gordon, J. M., Pascarelle, S. M., Schmidtke, P. C.,
Keel, W. C., Burkey, J. M., \& Dunlop, J. S.\ 1994, ApJ, 435, 577
\re
Windhorst, R. A., Keel, W. C., \& Pascarelle, S. M.\ 1998, ApJ, 494, L27
\re
Yamada, T., Ohta, K., Tomita, A., \& Takata, T.\ 1995, AJ, 110, 1564
\re
Zheng, W., Kriss, G. A., Telfer, R. C., Grimes, J. P., \& Davidsen,
A. F.\ 1997, ApJ, 475, 469 

\label{last}
\clearpage
\onecolumn

\begin{fv}{1}
{0cm}
{{\it JHK\/} spectra of 53W002. The lower trace shows the
1$\sigma$ noise level calculated from the background level. Many
strong forbidden lines and weak H$\beta$ line are present. Due to the
low-wavelength resolution, H$\alpha$+[N\,{\sc
ii}]\,$\lambda\lambda$6548,6583 and [S\,{\sc 
ii}]\,$\lambda\lambda$6716,6731 are blended.}
\bigskip
\begin{center}
\begin{minipage}{100mm}
\epsfxsize=95mm \epsfbox[30 160 570 570]{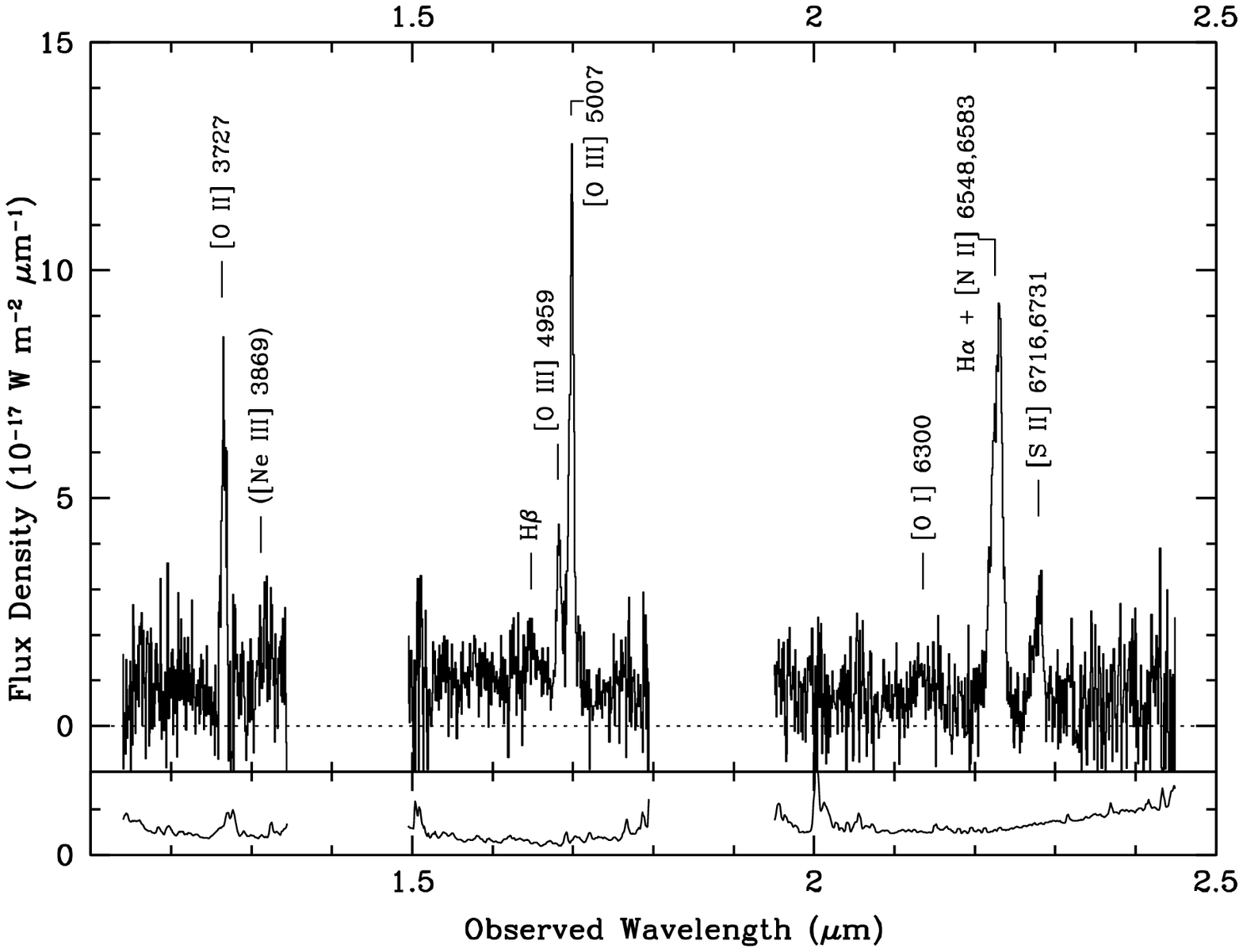}
\end{minipage}
\end{center}
\end{fv}

\begin{fv}{2}
{0cm}
{Deconvolution of the blended H$\alpha$+[N\,{\sc ii}] and
[S\,{\sc ii}] lines.}
\bigskip
\begin{center}
\begin{minipage}{75mm}
\epsfxsize=70mm \epsfbox{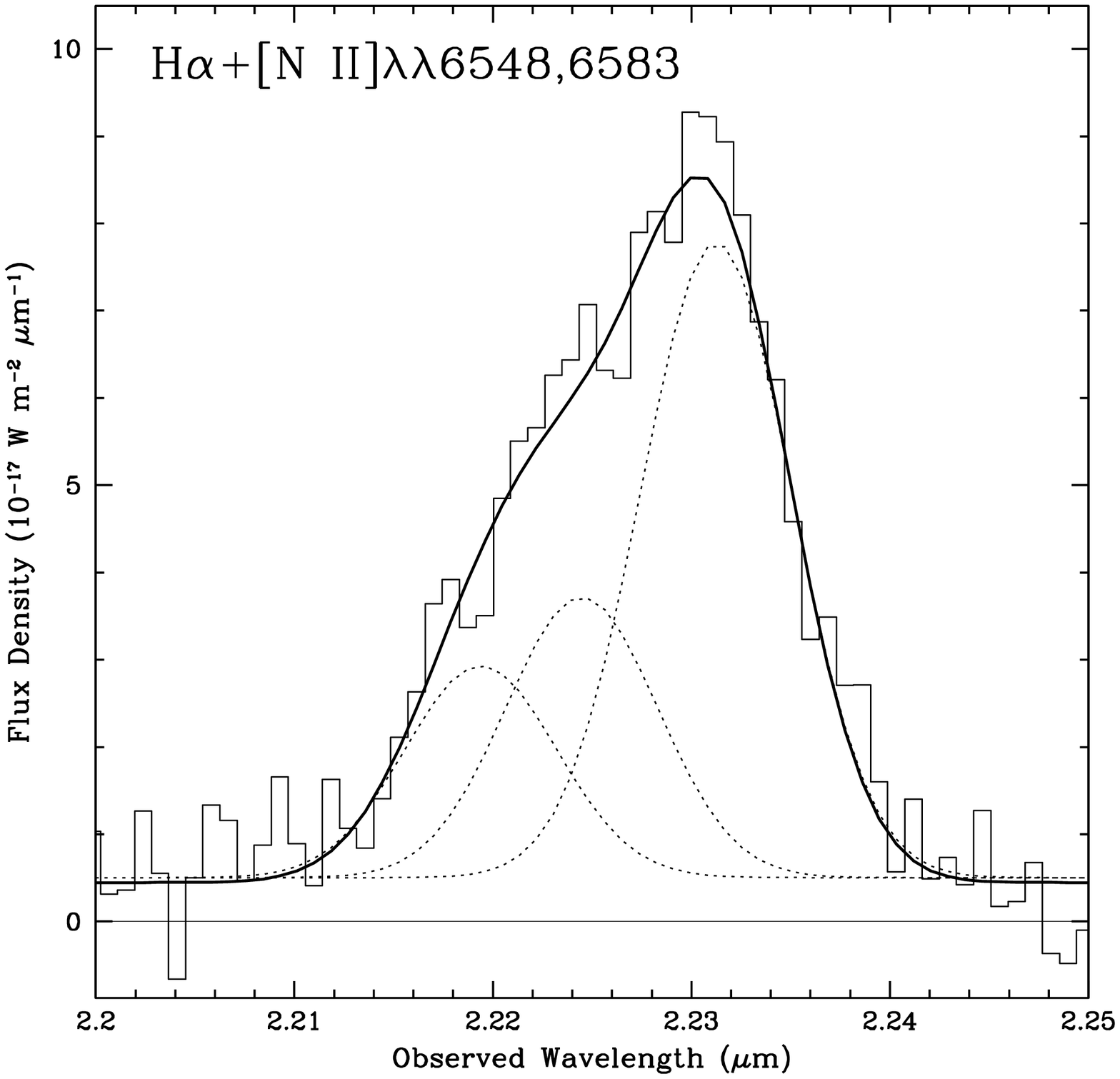}
\end{minipage}
\begin{minipage}{75mm}
\epsfxsize=70mm \epsfbox{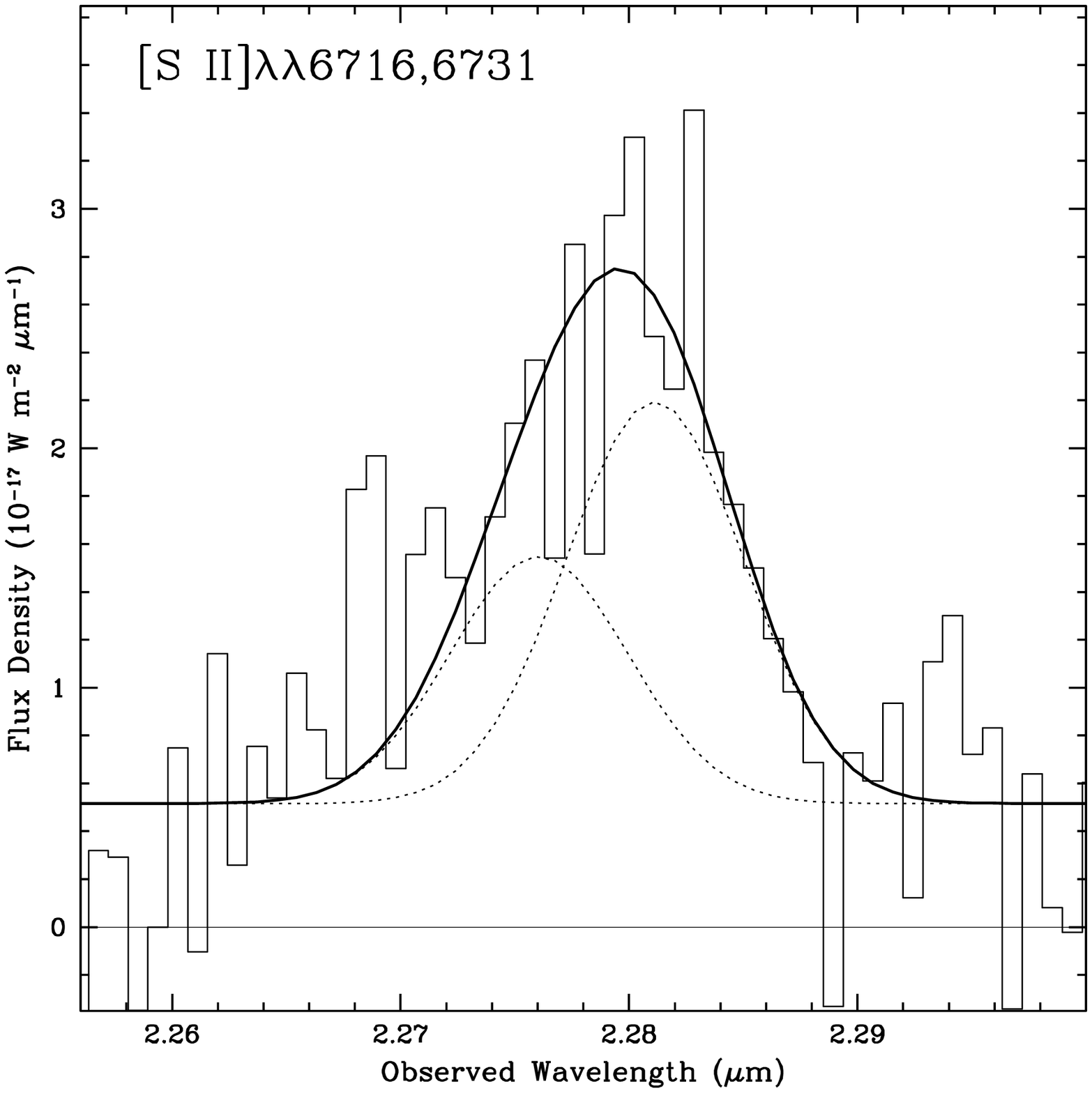}
\end{minipage}
\end{center}
\end{fv}

\begin{fv}{3}
{0cm}
{Spatial profiles of the spectra of 53W002. In the upper
panels, west is to the right, while south-east is to the right in the
lower two.  The dotted lines are fitted Gaussian profiles.  Both [O {\sc
iii}] and H$\alpha$+[N {\sc ii}] lines are more spatially extended
than the continuum. The seeing sizes for the $H$-band and $K$-band
spectra are $0.\!\!^{\prime\prime}6$ and $0.\!\!^{\prime\prime}4$,
respectively.
}
\bigskip
\begin{center}
\begin{minipage}{90mm}
\epsfxsize=85mm \epsfbox{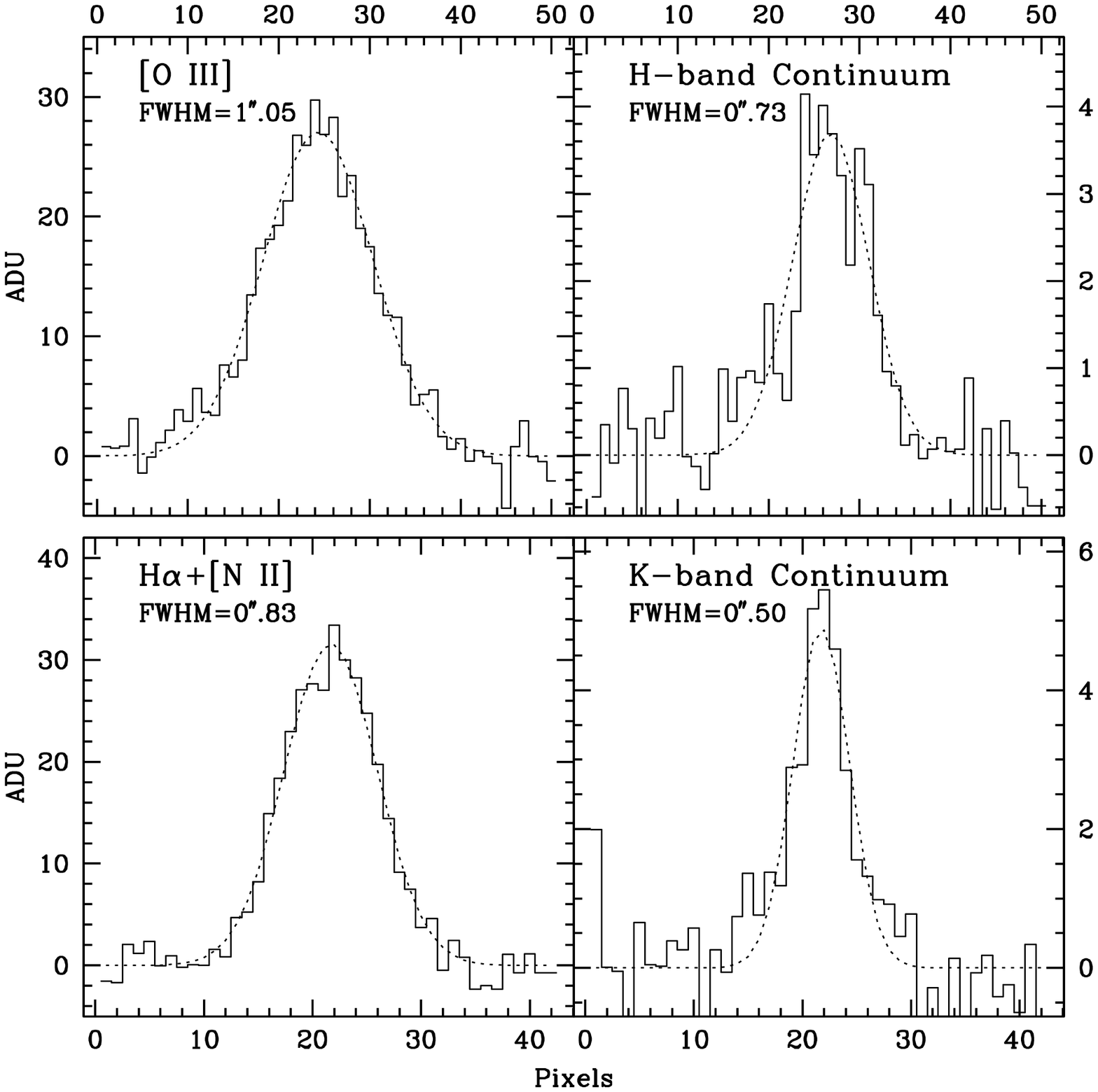}
\end{minipage}
\end{center}
\end{fv}

\begin{fv}{4}
{0cm}
{Continuum of 53W002. All of the emission lines, except
[Ne\,{\sc iii}], have been subtracted by Gaussian fitting, and the
resulting spectrum has been smoothed with a $0.05\, \mu \rm m$ boxcar
filter. The solid line shows the raw continuum, while the dotted line
shows further subtraction of [Ne\,{\sc iii}], where we have
assumed [Ne\,{\sc iii}]/[O\,{\sc ii}]\ =\ 0.2 (see text).}
\bigskip
\begin{center}
\begin{minipage}{90mm}
\epsfxsize=85mm \epsfbox{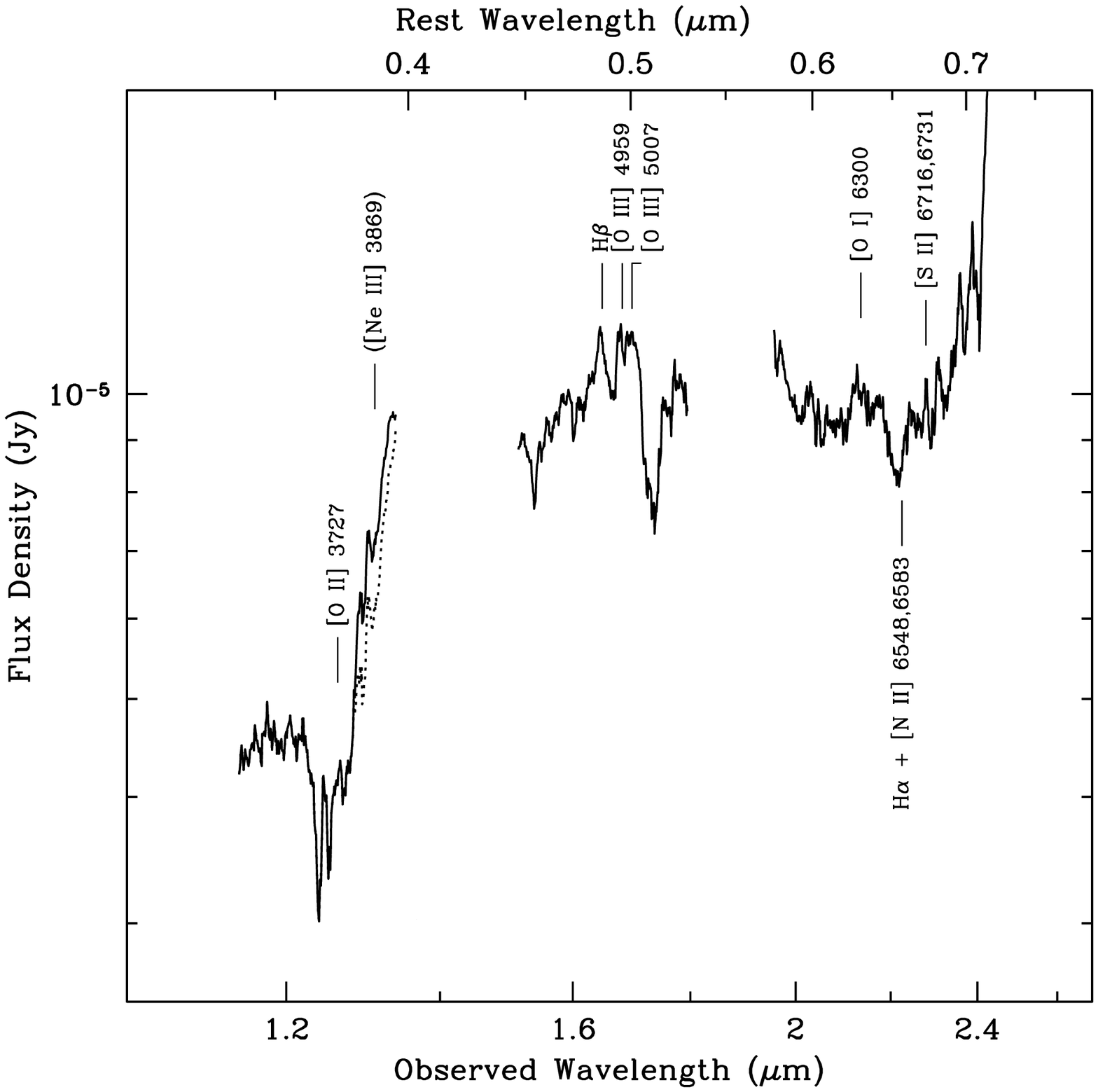}
\end{minipage}
\end{center}
\end{fv}

\begin{fv}{5}
{0cm}
{Reddening corrected line ratios of 53W002 (open circle with
error bars) plotted together with those of Seyfert 2s (filled
circles), radio galaxies (filled squares), H\,{\sc ii} regions (open
circles), H\,{\sc ii} galaxies (open circles), and starburst galaxies
(open triangles) taken from the literature (Veilleux, Osterbrock
1987; McCall et al.\ 1985; French 1980; Koski 1978; Costero,
Osterbrock 1977). The results of photoionization calculations carried
out using CLOUDY94 are indicated by the lines. Solar metallicity and a
spectral index of $\alpha=-0.7$ for the ionizing continuum were
assumed, except for one case where we assumed $\alpha=-1.5$ (thin
dot-dashed line), and one case of $Z=0.1 Z_{\odot}$ (thin
dash-long-dashed line).  The ionization parameter ($U$) varies along
each line, with representative points labeled by the value of $\log
U$.}
\bigskip
\begin{center}
\begin{minipage}{75mm}
\epsfxsize=70mm \epsfbox{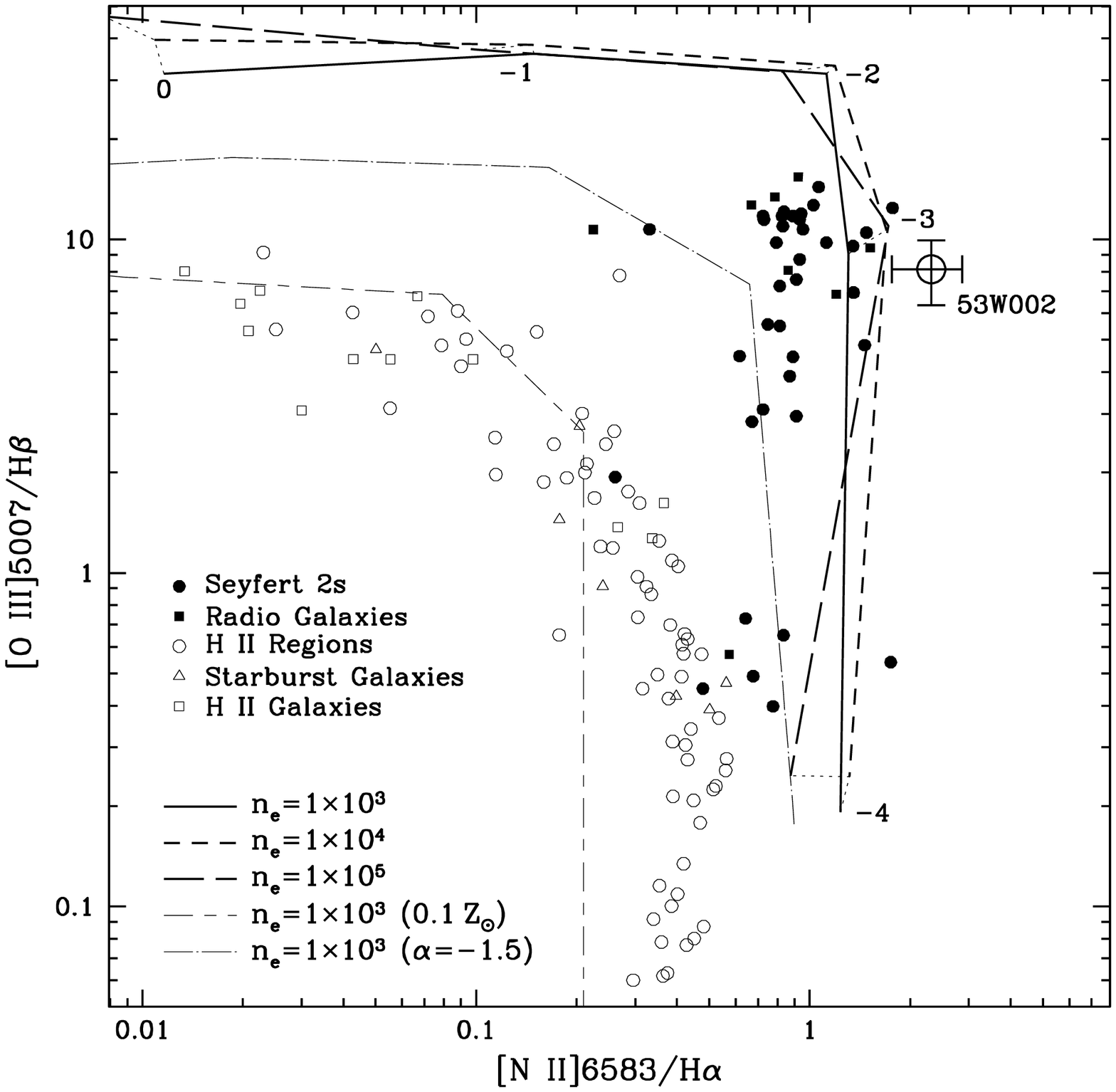}
\end{minipage}
\begin{minipage}{75mm}
\epsfxsize=70mm \epsfbox{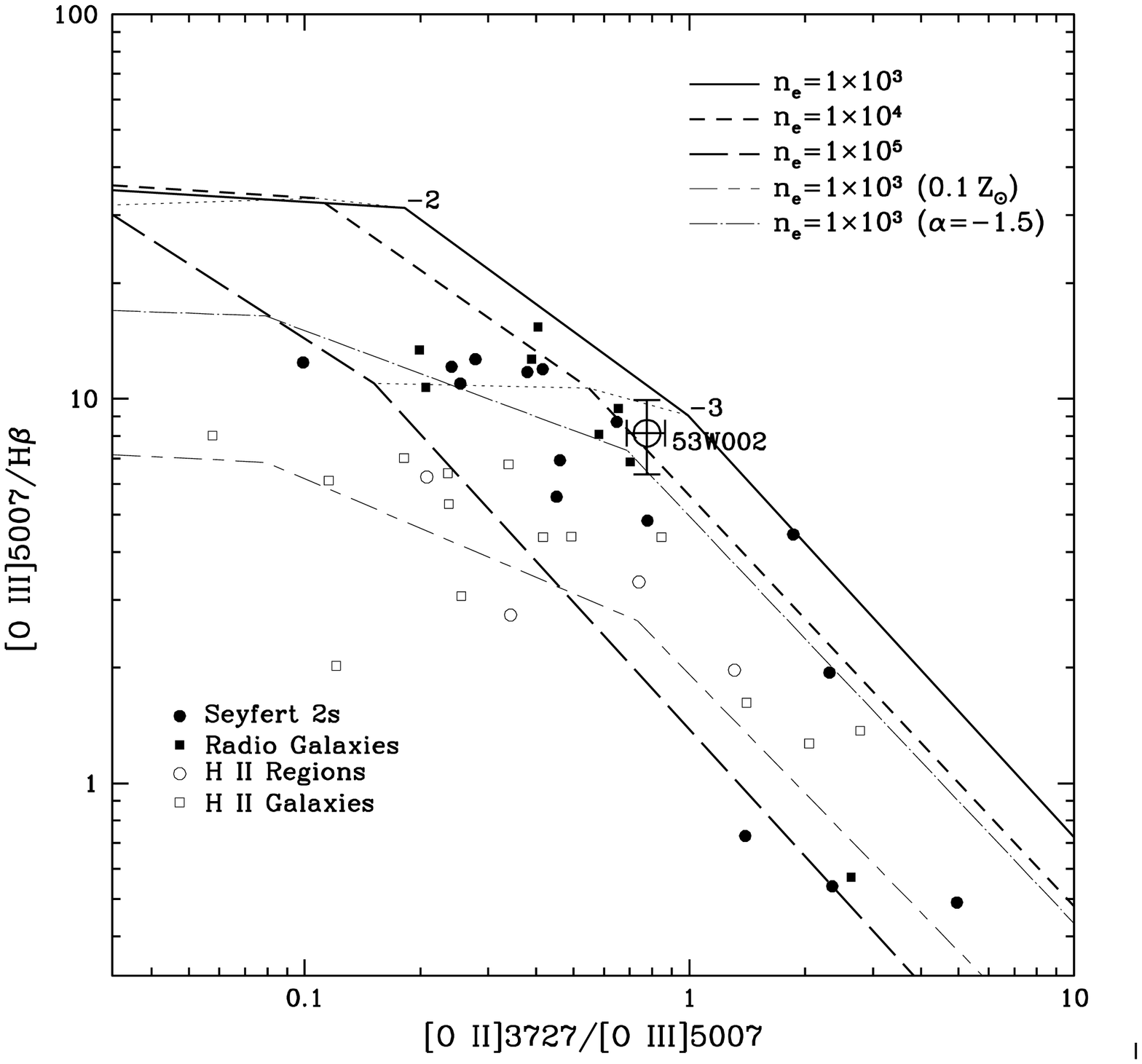}
\end{minipage}
\end{center}
\end{fv}

\begin{fv}{6}
{0cm}
{Spectral energy distributions of 53W002. Open circles and
triangles are observed values from the present observations and
Windhorst et al.\ (1991), respectively. The assumed AGN and nebular
continuum components were subtracted (see text) to produce the
SED of the stellar component alone, which is shown by the filled
symbols. The filled squares indicate the upper limit of the unresolved AGN
component determined from HST observations (Windhorst et al.\ 1998).
The nebular continuum was calculated from the H$\beta$ flux, and is shown
by the long dashed line. The solid lines show the best-fit
one-component spectral model. The dashed line shows the AGN
contribution, which is assumed to be dust-scattered in the left panel,
and electron-scattered in the right panel.}
\bigskip
\begin{center}
\begin{minipage}{75mm}
\epsfxsize=70mm \epsfbox{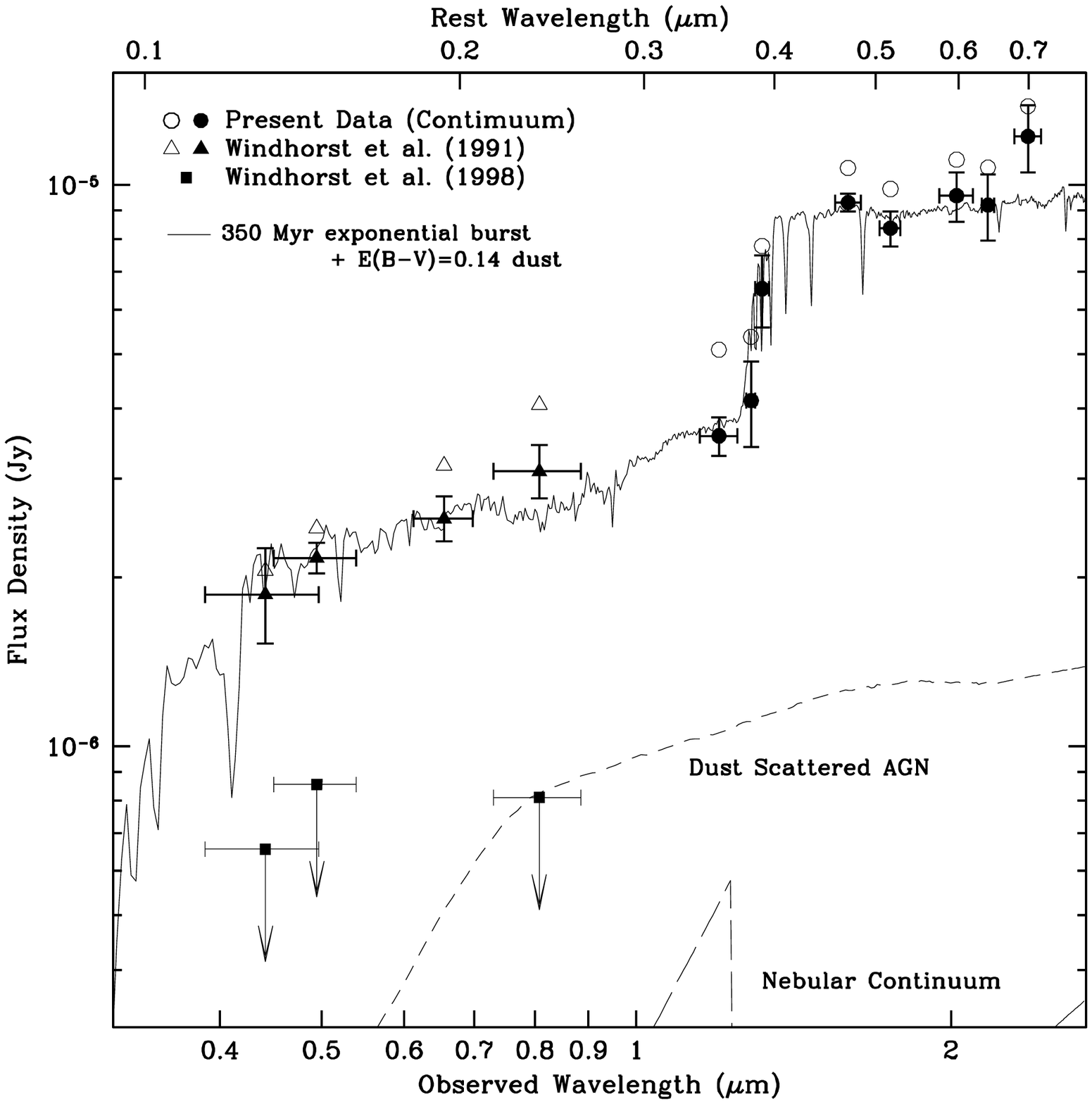}
\end{minipage}
\begin{minipage}{75mm}
\epsfxsize=70mm \epsfbox{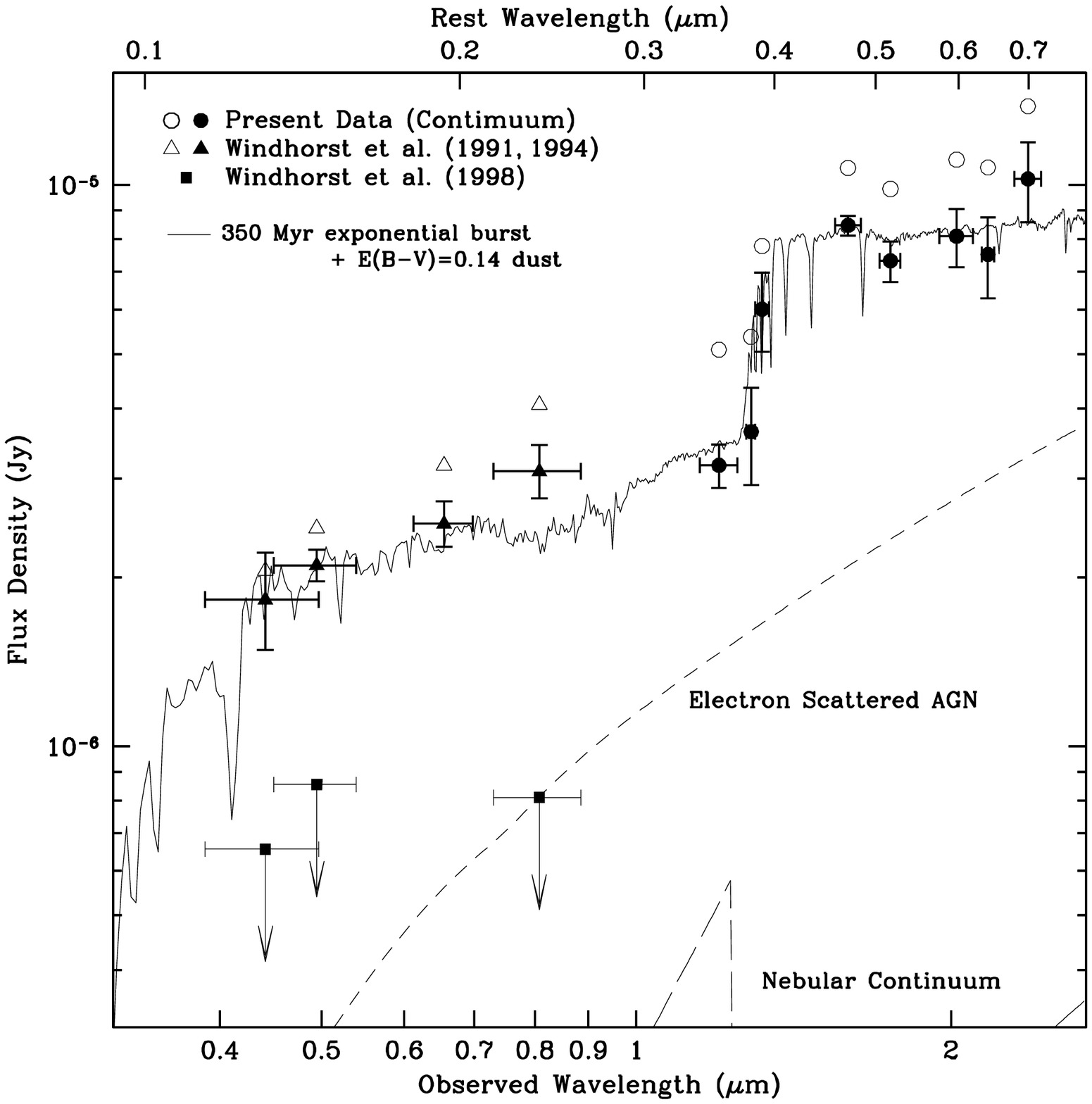}
\end{minipage}
\end{center}
\end{fv}

\begin{fv}{7}
{0cm}
{Results of fitting two-component models to the stellar SED of
53W002. The AGN component is assumed to be a dust-scattered power-law.
The dotted lines show the spectra of the old component, while the dashed
lines show that of the current burst.}
\bigskip
\begin{center}
\begin{minipage}{75mm}
\epsfxsize=70mm \epsfbox{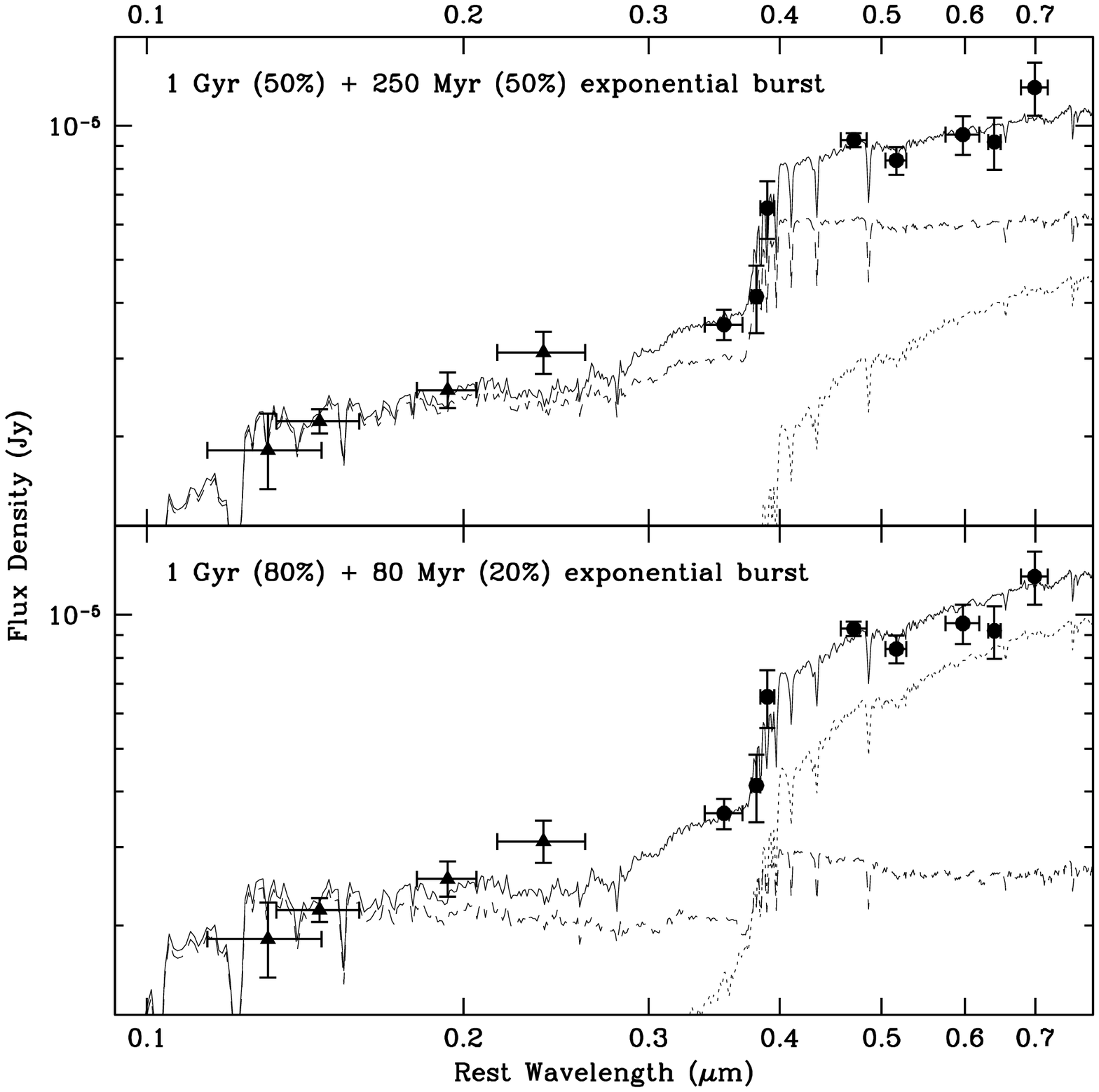}
\end{minipage}
\begin{minipage}{75mm}
\epsfxsize=70mm \epsfbox{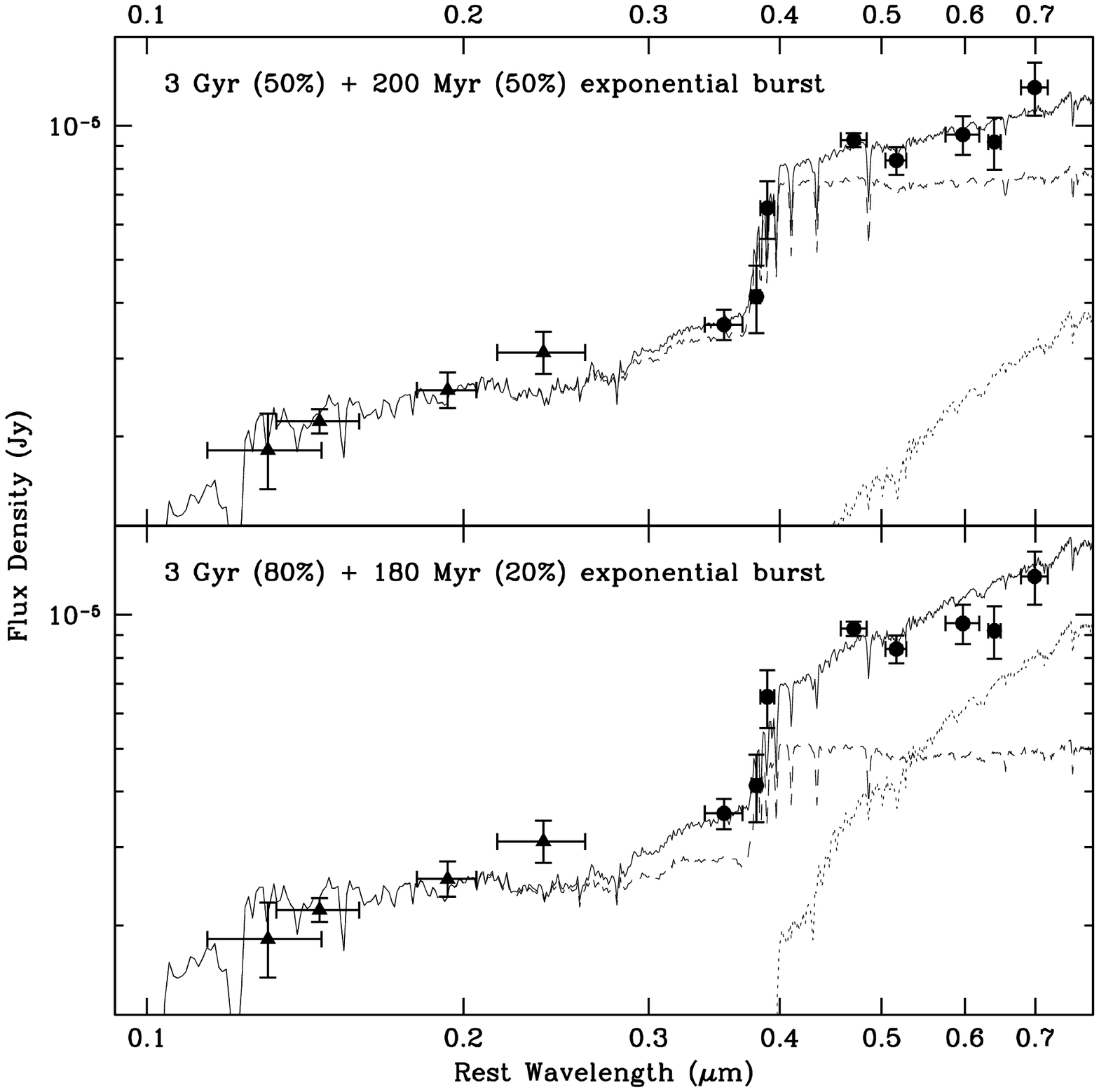}
\end{minipage}
\end{center}
\end{fv}

\clearpage

\begin{table*}
\small
\begin{center}
Table~1.\hspace{4pt}Emission line properties of 53W002. All errors are 2$\sigma$.\\
\end{center}
\vspace{6pt}

\begin{tabular}{ccccccc}
\hline
\hline
Line & $\lambda_{\rm rest}$& Redshift$^{*}$ & Flux & Flux (corrected) & Continuum & $EW_{\rm rest}$\\
& (\AA) && ($\times 10^{-19} \rm\,W\,m^{-2}$) & ($\times 10^{-19} \rm\,W\,m^{-2}$) &($\times 10^{-18} \rm \,W\,m^{-2}\,\mu m$) & (\AA)\\
\hline
{[}O\,{\sc ii}{]} & 3727 &2.3923$\pm$0.0004& 4.8$\pm$0.4 & 8.8$\pm$0.8 & 6.2$\pm$1.3 & 230$\pm$50\\
H$\beta$ & 4861 &2.3923$\pm$0.0004& 0.88$\pm$0.18 &1.4$\pm$0.3& 10.3$\pm$0.8 & 20$\pm$8 \\
{[}O\,{\sc iii}{]} & 4959 &2.3923$\pm$0.0004& 2.1$\pm$0.3 &3.3$\pm$0.4& 10.3$\pm$0.9 & 61$\pm$9\\
{[}O\,{\sc iii}{]} & 5007 &2.3923$\pm$0.0004& 7.3$\pm$0.5 &11.3$\pm$0.8& 10.3$\pm$0.9 & 210$\pm$20\\
{[}O\,{\sc i}{]}& 6300 &2.3833$\pm$0.0076& 0.76$\pm$0.36 &1.1$\pm$0.5& 5.9$\pm$1.4 & 37$\pm$20\\
H$\alpha$ & 6563 &2.3895$\pm$0.0008& 3.1$\pm$0.7 &4.3$\pm$0.9& 5.0$\pm$1.6 & 180$\pm$70\\
{[}N\,{\sc ii}{]} & 6583 &2.3895$\pm$0.0008& 7.2$\pm$0.8 &9.9$\pm$1.1& 5.0$\pm$1.6 & 420$\pm$140 \\
{[}S\,{\sc ii}{]} & 6716 &2.3890$\pm$0.0032& 1.0$\pm$0.7 &1.4$\pm$1.0& 5.2$\pm$2.5 & 60$\pm$60 \\
{[}S\,{\sc ii}{]} & 6731 &2.3890$\pm$0.0032& 1.6$\pm$0.7 &2.2$\pm$1.0& 5.2$\pm$2.5 & 90$\pm$70\\
\hline
\end{tabular}
\vspace{6pt}\par\noindent
 *~ No systematic errors are included.
\end{table*}

\begin{table*}
\small
\begin{center}
Table~2.\hspace{4pt} Parameters for the best-fit models of the spectral energy
 distribution.\\
\end{center}
\vspace{6pt}

\begin{tabular}{p{13pc}ccccc}
\hline
\hline
\multicolumn{1}{c}{Galaxy evolution model} & Non-stellar$^*$ & $\chi^2$& Age (Myr)
 & Stellar mass ($M_{\odot}$)& SFR ($M_{\odot}\,\rm yr^{-1}$)\\
\hline
Instantaneous burst\dotfill 
&N& 19.1 & 80 &$9.7\times10^{10}$ &  0 \\
&E& 19.0 & 70 &$6.8\times10^{10}$ &  0 \\
&D& 19.1 & 80 &$8.1\times10^{10}$ &  0 \\
Exponential burst ($\tau=200$ Myr)\dotfill 
&N& 21.3 & 350 &$1.3\times10^{11}$ & $1.5\times10^{2}$ \\
&E& 12.5 & 350 &$9.7\times10^{10}$ & $1.1\times10^{2}$ \\
&D& 10.0 & 350 &$1.1\times10^{11}$ & $1.2\times10^{2}$ \\
Exponential burst ($\tau=1$ Gyr)\dotfill 
&N& 23.8 & 600 &$1.4\times10^{11}$ & $1.8\times10^{2}$ \\
&E& 11.0 & 600 &$1.1\times10^{11}$ & $1.4\times10^{2}$ \\
&D& 8.9 & 700 &$1.3\times10^{11}$ & $1.4\times10^{2}$ \\
\hline
\end{tabular}
\vspace{6pt}\par\noindent
 *~ Model of non-stellar components. N: No non-stellar component is
 assumed. D: Nebular continuum + dust scattered $\alpha=-0.7$ power
 law. E: Nebular continuum + electron scattered $\alpha=-0.7$ power law. 
\end{table*}

\begin{table*}
\small
\begin{center}
Table~3.\hspace{4pt}  Best-fit two-components models of the spectral energy
 distribution.\\
\end{center}
\vspace{6pt}

\begin{tabular}{p{5pc}cccccccc}
\hline
\hline
\multicolumn{2}{c}{Age of old component}&\multicolumn{3}{c}{1 Gyr}
&\multicolumn{3}{c}{3 Gyr}\\
\hline
&&&Age$^{\ddag}$ & Stellar mass&&Age$^{\ddag}$ & Stellar mass\\
\multicolumn{1}{c}{\raisebox{1.5ex}[0pt]{Non-stellar$^*$}} & 
\multicolumn{1}{c}{\raisebox{1.5ex}[0pt]{Old frac.$^{\dag}$}}& 
\multicolumn{1}{c}{\raisebox{1.5ex}[0pt]{ $\chi^2$}}& 
(Myr) &$(M_{\odot})$&
\multicolumn{1}{c}{\raisebox{1.5ex}[0pt]{ $\chi^2$}}& 
(Myr) &$(M_{\odot})$\\
\hline
D\dotfill
& 0.1 & 9.5 & 350 & 1.5$\times 10^{11}$ & 9.2 & 350 & 1.2$\times 10^{11}$\\
& 0.2 & 11.4 & 350 & 1.6$\times 10^{11}$ & 9.0 & 350 & 1.3$\times 10^{11}$\\
& 0.3 & 8.9 & 300 & 1.6$\times 10^{11}$ & 9.9 & 350 & 1.5$\times 10^{11}$\\
& 0.4 & 10.0 & 300 & 1.7$\times 10^{11}$ & 9.6 & 300 & 1.5$\times 10^{11}$\\
& 0.5 & 8.6 & 250 & 1.8$\times 10^{11}$ & 8.4 & 300 & 1.7$\times 10^{11}$\\
& 0.6 & 9.5 & 200 & 1.9$\times 10^{11}$ & 11.8 & 300 & 2.1$\times 10^{11}$\\
& 0.7 & 9.8 & 160 & 2.1$\times 10^{11}$ & 12.2 & 250 & 2.3$\times 10^{11}$\\
& 0.8 & 13.3 & 80 & 2.4$\times 10^{11}$ & 18.7 & 180 & 2.7$\times 10^{11}$\\
& 0.9 & 84.4 & 40 & 2.9$\times 10^{11}$ & 45.8 & 70 & 3.7$\times 10^{11}$\\
\hline
E\dotfill
& 0.1 & 12.4 & 300 & 1.3$\times 10^{11}$ & 13.5 & 300 & 9.3$\times 10^{10}$\\
& 0.2 & 11.2 & 300 & 1.4$\times 10^{11}$ & 12.1 & 300 & 1.0$\times 10^{11}$\\
& 0.3 & 12.8 & 300 & 1.5$\times 10^{11}$ & 11.3 & 300 & 1.2$\times 10^{11}$\\
& 0.4 & 11.7 & 250 & 1.5$\times 10^{11}$ & 11.7 & 300 & 1.4$\times 10^{11}$\\
& 0.5 & 14.2 & 250 & 1.6$\times 10^{11}$ & 14.6 & 300 & 1.6$\times 10^{11}$\\
& 0.6 & 13.1 & 180 & 1.7$\times 10^{11}$ & 13.8 & 250 & 1.7$\times 10^{11}$\\
& 0.7 & 14.6 & 140 & 1.9$\times 10^{11}$ & 18.1 & 200 & 1.9$\times 10^{11}$\\
& 0.8 & 19.5 & 50 & 2.2$\times 10^{11}$ & 24.4 & 160 & 2.4$\times 10^{11}$\\
& 0.9 & 103.4 & 40 & 2.6$\times 10^{11}$ & 54.5 & 60 & 3.4$\times 10^{11}$\\
\hline
\end{tabular}
\vspace{6pt}
\par\noindent
 *~ Model of non-stellar components. D: Nebular continuum + dust scattered $\alpha=-0.7$ power
 law. E: Nebular continuum + electron scattered $\alpha=-0.7$ power
 law. 
\par\noindent
 $\dag$~ Mass fraction of old component.
\par\noindent
$\ddag$ Age of current burst component.
\end{table*}

\end{document}